\documentclass[12pt,a4paper]{article}

\usepackage{amsmath}
\usepackage{amssymb}
\usepackage{amsfonts}

\usepackage{eqsection}

\usepackage{hyperref}

\usepackage{graphicx}

\newcommand{\de}{\partial}

\newcommand{\calD}{{\mathcal D}_n}
\newcommand{\calF}{\mathcal F}
\newcommand{\calH}{\mathcal H}

\newcommand{\calU}{\mathcal U}
\newcommand{\calZ}{\mathcal Z}

\newcommand{\J}{\mathrm J}

\newcommand{\<}{\langle}
\renewcommand{\>}{\rangle}

\newcommand{\osp}{\mathop{\rm osp}\nolimits}
\newcommand{\ssp}{\mathop{\rm sp}\nolimits}

\renewcommand{\Re}{\mathop{\rm Re}\nolimits}

\newtheorem{lemma}{Lemma}
\date{June 24, 2009}

\title{Phase transition in the spanning-hyperforest model 
on complete hypergraphs}
\author{Andrea Bedini, Sergio Caracciolo and 
  Andrea Sportiello \\
  {\scriptsize Dipartimento di Fisica dell'Universit\`a degli Studi di
    Milano and INFN, Sezione di Milano,} \\[-2mm]
  {\scriptsize via Celoria 16, I-20133 Milano, Italy}\\
  {\tt \scriptsize Andrea.Bedini@mi.infn.it,
    Sergio.Caracciolo@mi.infn.it,
    Andrea.Sportiello@mi.infn.it}}

\begin{document}

\maketitle

\abstract{
By using our novel Grassmann formulation we study the phase transition
of the spanning-hyperforest model of the $k$-uniform complete
hypergraph for any $k\geq 2$. The case $k=2$ reduces to the
spanning-forest model on the complete graph. Different $k$ are studied
at once by using a microcanonical ensemble in which the number of
hyperforests is fixed. The low-temperature phase is characterized by
the appearance of a giant hyperforest. The phase transition occurs
when the number of hyperforests is a fraction $(k-1)/k$ of the total
number of vertices. The  behaviour at criticality is also studied by
means of the coalescence of two saddle points. As the Grassmann
formulation exhibits a global supersymmetry we show that the phase
transition is second order and is associated to supersymmetry breaking
and we explore the pure thermodynamical phase at low temperature by
introducing an explicit breaking field.
}

\clearpage

\section{Introduction}

The phase transition in a model of spanning forests is particularly interesting because only the geometric properties of connection of different parts are involved and this extremely reduced structure is probably at the root of many critical phenomena, within, and even outside, natural sciences.

A possible way to attack this problem on a generic graph by the tools of statistical mechanics goes back to the formulation as a Potts model~\cite{Potts, Wu, Wu2} in the limit of vanishing number  $q$ of states.
The Potts model on any finite graph $G=(V,E)$, with vertex set $V$ and edge set $E$, is characterized by the  coupling $v_e$, for each edge $e\in E$, which is related to the inverse temperature $\beta$ and exchange coupling $J_e$ by the relation $v_e = e^{\beta J_e} - 1$. By definition $q$
is a positive integer and the set of couplings ${\bf v} = \{v_e\}_{e\in E}$ is of real numbers in the interval $[-1,\infty]$. The Fortuin-Kasteleyn representation~\cite{Kasteleyn_69, Fortuin_72} expresses the partition function $Z_G(q; {\bf v})$ of the Potts model as a sum on all subgraphs $H\subseteq G$ of monomials in both $q$ and  $v_e$'s 
\begin{equation}
Z_G(q; {\bf v}) := \sum_{H\subseteq G} q^{K(H)}\,\prod_{e\in E(H)} v_e
\end{equation}
where $K(H)$ is the number of connected components of the subgraph $H$. Therefore the model is easily extended to more general values of its parameters. In this form it takes the name of {\em random cluster model}~\cite{Grimmet}.
More generally, it is convenient to introduce the redundant description in terms of two global parameters $\lambda$ and $\rho$ 
\begin{equation}
Z_G(\lambda,\rho; {\bf w}) := \sum_{H\subseteq G} \lambda^{K(H)-K(G)}\,\rho^{L(H)}\, \prod_{e\in E(H)} w_e
\end{equation}
where $L(H)$ is the {\em cyclomatic number} of the subgraph $H$. The redundancy is easily shown by using the Euler relation
\begin{equation}
V -  K = E - L
\end{equation}
and the relations
\begin{align}
q \,=\, & \lambda \, \rho \\
{\bf v} \,=\, & {\bf w} \, \rho \, .
\end{align}
Indeed, this form is suitable for taking two different limits when $q \to 0$. In the former limit $\lambda\to 0$ at $\rho$ fixed  only maximally-connected subgraphs will survive and will be weighted by a factor $\rho^{L(H)}$. In the latter one $\rho\to 0$ at $\lambda$ and $\bf w$ fixed only spanning forests will survive, weighted by a factor $\lambda^{K(H)}$ as $\lambda^{-K(G)}$ is an overall constant. Remark that, when $G$ is {\em planar}, maximally-connected subgraphs are in one-to-one correspondence with spanning forests by graph duality, and that when both $\lambda\to 0$ and $\rho\to 0$ only spanning trees will survive in each connected component of $G$.

As shown in \cite{CarJacSal04, CarSokSpo07}, 
the model in the limit $\rho\to 0$, that is the spanning forest model, admits a representation in terms of fermionic
fields, which means that the partition function can be written as a multiple
Berezin integral over anti-commuting variables which belong to a Grassmann algebra. Moreover, this
representation~\cite{CarSokSpo07, BedCarSpo08} is powerful enough to describe the model with many body interactions which gives rise to hyperforests defined on a
hypergraph~\cite{Grimmett_94}, which is a natural generalization of the concept of graph where
the edges can connect more than two vertices at once.

In two dimensions, the critical behaviour of the ferromagnetic Potts/random-cluster
model is quite well understood, thanks to a combination of exact solutions~\cite{Bax07}, Coulomb gas methods~\cite{Nienhuis}, and conformal field theory~\cite{FraMatSen97}. 
Information can also be deduced from the study of the model on random planar lattices~\cite{Kazakov,PZJ,eynbon}. Also in the $q \to 0$ limit detailed results are avalaible both for the tree model, in particular in connection with the abelian sandpile model~\cite{MD},
as for spanning forest on a regular lattice~\cite{CarJacSal04,claudia} and directly in the continuum~\cite{arboreal}. Also  the model 
on random planar lattices has been considered~\cite{BP}.

But in more than two dimensions the only quantitative informations we have about the
spanning-forest model come from numerical investigations~\cite{DenGarSok07}.  Monte Carlo simulations performed at increasing dimensionality ($d
= 3,4,5$) show a second-order phase transition.

Much less results are available for the case of hyperforests. Also in two dimensions or in the limit of hypertrees. 
Even the problem of determining whether there exists a spanning hypertree in a given $k$-uniform hypergraph, 
is hard, technically NP-complete, for $k\ge 4$, whereas for $k=3$, there exists a polynomial-time algorithm based on Lovasz' theory of polymatroid matching~\cite{LovPlu}. See~\cite{CMSS} for a randomized polynomial-time algorithm in the case $k=3$ whose main ingredients is a Pfaffian formula  for a polynomial that enumerates spanning hypertrees with some signs~\cite{MV}, which is quite similar to our Grassmann representation~\cite{Abdessalam}. 

In~\cite{SPS} a phase transition is detected in the random $k$-uniform hypergraph when a number of hyperedges $|E|=n/k(k-1)$ of the total number of vertices $n=|V|$ is chosen uniformly at random.   
In the case of random graphs, that is for $k=2$,  
Erd\H{o}s and R\'enyi showed in their classical paper~\cite{ER}  that at the transition an abrupt change occurs in the structure of the graph, for low density of edges it consists of many small components, while, in the high-density regime a {\em giant} component occupies  a finite fraction of the vertices. 
Remark that their ensemble of subgraphs is the one occurring in the microcanonical formulation, at fixed number of edges,  of the Potts model at number of states $q=1$.
The connected-component structure of the random $k$-uniform hypergraph has been analyzed in~\cite{SPS} where it has been shown that if $|E| < n/k(k-1)$ the largest component occupies order $\log n$ of vertices,  for $|E| = n/k(k-1)$ it has order $n^{2/3}$ vertices and for $|E| > n/k(k-1)$ there is a unique component with order $n$ vertices. More detailed information on the behaviour near the phase transition when $|E| \to  n/k(k-1)$
have been recovered in~\cite{Bollo1, Bollo2} for the case of the random graph, but see also~\cite{JKLP, JLR},  and in~\cite{Karonski_02} for the general case of hypergraphs.

By using the new Grassmann representation,  we present here a study of the phase transition for the hyperforest model on the $k$-uniform complete hypergraph, for general $k$, where the case $k=2$ corresponds to spanning forests on the complete graph. The random-cluster model~\cite{BolGriJan05} on the complete graph has already been developed but it cannot be extended to the $q \to 0$ case, exactly like the mean-field solution for the Potts model~\cite{Mittag, Baracca}. The fermionic representation, instead, describes the Potts model directly at $q=0$ as it provides an exact representation of the partition function of the spanning-hyperforest model.

As usual with models on the complete graph, the statistical weight reduces to a
function of only one extensive observable, which here is quadratic in the Grassmann variables. Under such a condition  the partition function can be expressed as the integration over a single complex variable in a closed contour around the origin~\cite{BedCarSpo08}.
Indeed, counting the spanning forests over a complete (hyper-)graph is indeed a typical problem of {\em analytical combinatorics}.  And, exactly like in the case of ordinary graph, when the number of connected components in the spanning forests is macroscopic, that is a finite fraction of the number of vertices, there are two different regimes, which can be well understood by means of two different saddle points of a closed contour integration over a single complex variable as presented in \cite{FlaSed08} (but see also the probabilistic analysis in~\cite{Kolchin}). And even the behaviour at the critical point can be studied as the coalescence of these two saddle points.

In this paper we shall first review for reader convenience the Grassmann formulation of the spanning-forest model in Sec.~\ref{sec:the_spanning_forests_model} and in Sec.~\ref{sec:the_mean_field_theory} how it is possible to recover, in the case of the $k$-uniform complete hypergraphs,  a representation of the partition function suitable for the asymptotic analysis for large  number of vertices $n$. In this same Section we shall also present a full discussion of the saddle points in the micro-canonical ensemble, that is at fixed number of connected components, and of the associated different phases.  We shall see that the universality class of the transition is independent from $k$. And we will exhibit the relation with the canonical ensemble.
In Sec.~\ref{sec:t} we will provide an interpretation of the transition as the appearance of a {\em giant} component by introducing a suitable observable which is sensible to the size of the different hypertrees in the hyperforest.

More interestingly, our Grassmann formulation exhibits a global continuos supersymmetry, non-linearly realized.
We shall show that the phase transition is associated to the spontaneous breaking of this supersymmetry.
By the introduction of an explicit breaking we shall be able to investigate the expectation values in the broken pure thermodynamical states. We shall therefore be able to see in Sec.~\ref{sec:the_symmetry_breaking} that the phase transition is of second order. This seems at variance with the supersymmetric formulation of polymers given by Parisi and Sourlas~\cite{Parisi-Sourlas} where it appeared to be of zeroth order. 

\section{The spanning-forest model} 
\label{sec:the_spanning_forests_model}

Given the complete hypergraph $\overline{\cal K}_n=(V,E)$ with vertex set $V = [n]$, and complete in the $k$-hyperedges for all $2\leq k\leq n$,  so that the hyperedge set $E$ is the collection of all $A \subseteq V$ with cardinality at least 2, let's introduce on each vertex $i \in V$ a pair of anti--commuting variables
$\psi_i$, $\bar\psi_i$:
\begin{equation}
  \{\psi_i, \psi_j\} = \{\bar\psi_i, \bar\psi_j\} = \{\bar\psi_i, \psi_j\} = 0
  \qquad
  \forall i, j \in V(G)
\end{equation}
which generate the Grassmann algebra $\Lambda[\psi_1, \dots, \psi_n,\bar\psi_1, \dots, \bar\psi_n]$ of dimension $2^{2n}$.

Then, for each hyperedge $A\subseteq E$, we define the monomial
\begin{equation}
	\tau_A := \prod_{i \in A} \bar\psi_i \psi_i
	,
\end{equation}
and, for each indeterminate $t$, the Grassmann element
\begin{equation}
  f_A^{(t)} := t (1 - |A|) \tau_A + \sum_{i \in A} \tau_{A
    \smallsetminus i} - \sum_{\substack{i,j \in A \\
      i \neq j}} \bar\psi_i \psi_j \tau_{A \smallsetminus \{i,j\}}\, .
\end{equation}

In \cite{CarSokSpo07} it has been shown that the generating function of unrooted
spanning forest on a generic hypergraph admits the following representations:
given a set of edge weights $\mathbf w = \{w_{A}\}_{A \in E}$ we have
\begin{align}
  \calZ_G(\mathbf w, t) & := \sum_{F\in \calF} t^{K(F)} \prod\limits_{A \in F} w_A \\
	& \label{eq:z}
  = \int \calD (\bar\psi, \psi)\
	\exp \left\{
		t \sum_{i \in V} \bar\psi_i \psi_i + \sum_{A \in E} w_A f_A^{(t)}
	\right\}
	\\
  &  =  \int \calD (\bar\psi, \psi)\ \exp \left( - {\cal H} \right)
\end{align}
where
the indeterminate $t$ plays the role of the parameter $\lambda$ we had in the random cluster model formulation, $\cal F$ is the set of hyperforests, $K(F)$ the number of connected components in the hyperforest $F$, that  is the number of hypertrees, 
\begin{align}
	\int \calD (\bar\psi, \psi) := \int \prod_{i \in V} d\bar\psi_i d\psi_i
\end{align}
is the Berezin integration and we denoted by $- {\cal H}$, as usual in statistical mechanics, the exponential weight.


The fermionic model we
introduced above presents a non-linearly realized $\osp(1|2)$ supersymmetry~\cite{CarJacSal04, CarSokSpo07}. 
Firstly, we have the elements of the $\ssp(2)$ subalgebra,
 with
\begin{align}
 \delta \psi_i & = 
   -\, \alpha\, \psi_i + \gamma\,  \bar\psi_i     \\
 \delta  \bar\psi_i & = 
  +\, \alpha\, \bar\psi_i + \beta\, \psi_i
 \label{def.sp2}
\end{align}
where $\alpha,\beta,\gamma$ are bosonic (Grassmann-even) global parameters.
Secondly, we have the transformations parametrized by
fermionic (Grass\-mann-odd) global parameters $\epsilon,\bar\epsilon$:
\begin{align}
 \delta \psi_i & = 
   t^{-1/2}\,\epsilon \,(1 - t\,\bar\psi_i \psi_i)   
   \label{eq:osp-fermionica}
\\
 \delta \bar\psi_i & = 
    t^{-1/2}\,\bar\epsilon \,(1 - t\,\bar\psi_i \psi_i)
 \label{eq:osp-fermionicb}
\end{align}
In terms of the differential operators
$\partial_i = \partial/\partial \psi_i$
and $\bar{\partial}_i = \partial/\partial \bar\psi_i$,
the transformations~(\ref{def.sp2}) can be represented by the generators
\begin{align}
 X_0  & =  \sum_{i\in V} (\bar\psi_i \bar{\partial}_i - \psi_i \partial_i)
   \\[1mm]
 X_+  & =  \sum_{i\in V} \bar\psi_i \partial_i
   \\[1mm]
 X_-  & =  \sum_{i\in V} \psi_i \bar{\partial}_i
 \label{eq:repsp2}
\end{align}
corresponding to the parameters $\alpha,\beta,\gamma$, respectively,
while the transformations~(\ref{eq:osp-fermionica}) and~(\ref{eq:osp-fermionicb}) 
can be represented by the generators
\begin{align}
  Q_+  & = 
      t^{-1/2} \sum_{i\in V}  (1 - t \,\bar\psi_i \psi_i) \partial_i 
   \\[1mm]
  Q_-  & =   t^{-1/2} \sum_{i\in V}  (1 - t\, \bar\psi_i \psi_i)
      \bar{\partial}_i 
 \label{eq:def_Q_bis}
\end{align}
corresponding to the parameters $\epsilon,\bar\epsilon$, respectively.
These transformations satisfy the commutation/anticommutation relations
\begin{align}
   [X_0, X_\pm] \,=\, \pm 2 X_\pm
        \quad   &   \quad
   [X_+, X_-] \,=\, X_0
      \label{eq:sp2} \\[1mm]
   \{ Q_\pm, Q_\pm \} \,=\, \pm 2 X_\pm
        \quad   &   \quad
   \{ Q_+, Q_- \} \,=\,  X_0
      \label{eq:osp12a} \\[1mm]
   [X_0, Q_\pm] \,=\, \pm Q_\pm \quad\qquad
      [X_\pm, Q_\pm] &\!=\!  0  \quad\qquad
      [X_\pm, Q_\mp] \,=\,  -Q_\pm 
     \label{eq:osp12c}
\end{align}
Note in particular that $X_\pm = Q_\pm^2$ and $X_0 = Q_+ Q_- + Q_- Q_+$.

%





\section{Uniform complete hypergraphs} 
\label{sec:the_mean_field_theory}

The $k$-uniform complete hypergraph is the
hypergraph whose vertices are connected in groups of $k$ in all possible ways
or, alternatively, whose edge set $E$ is the set of all $k$-sets over the
vertex set $V$.  In our general formulas for the complete hypergraph we must set the weights $w_A=0$ for all the hyperedges $A$ with cardinality different from $|A|=k$.
In the following we will set all the nonzero weights to one, so that we shall restrict ourselves to a simple, one-parameter, counting problem. 
In this case the expansion of the partition function in series of $t$
\begin{equation}
\calZ(t) = \sum_p \calZ_p \,t^p
\end{equation}
provides $\calZ_p$ the number of hyperforests, with all hyperedges with cardinality $k$, composed by $p$ hypertrees.
Please remark that, in the $k$-uniform complete hypergraph,  the number $p$ of hyperforests must be such that 
\begin{equation}
s \, = \, \frac{n-p}{k-1} \label{s}
\end{equation}
must be an integer. Indeed it is the total number of hyperedges in the hyperforest.

By definition
\begin{align}
\calZ_p = & \, \frac{1}{p!}\,\langle {\cal U}^p \rangle_{t=0} \\
= &   \, \frac{1}{p!}\, \int \calD (\bar\psi, \psi)\, {\cal U}^p\,
	\exp \left\{ \sum_{A: |A|=k} f_A^{(0)}
	\right\}
\end{align}
where
\begin{equation}
{\cal U} =   \sum_{i \in V} \bar\psi_i \psi_i  +  (1 - k)  \sum_{A: |A|=k}  \tau_A  
\end{equation}
and $\langle \cdot \rangle_{t=0}$ is the un-normalized expectation value in the ensemble of hypertrees.

The interested reader can find a full comparison of our approach with respect to the standard tools of combinatorics in our previous paper~\cite{BedCarSpo08}. We introduce a mean-field variable
\begin{equation}
	\bar\psi\psi := \sum_{i \in V} \bar\psi_i \psi_i = (\bar\psi, \psi)
\end{equation}
and we observe that
\begin{align}
{\cal U} & = \bar\psi\psi + (1 - k) \frac {(\bar\psi\psi)^k} {k!}  \\
  \sum_{A: |A|=k} f_A^{(0)} & =
   n \frac{(\bar\psi\psi)^{k-1}}{(k-1)!} - (\bar\psi, \J \psi)
  \frac{(\bar\psi\psi)^{k-2}}{(k-2)!}
  ,
\end{align}
where $\J$ is the matrix with 1 on all entries, so that $(\bar\psi, \J \psi) = \sum_{i,j} \bar\psi_i \psi_j$. The following 
lemma then applies:
\begin{lemma}[\cite{BedCarSpo08}]
Let $n$ be the number of vertices, $g$ and $h$ generic functions on the Grassmann algebra, then
\begin{multline}
	\int \calD (\bar\psi, \psi)\ (\bar\psi\psi)^{r}
	e^{h(\bar\psi\psi) + (\bar\psi, J \psi) g(\bar\psi\psi)}
	\\
	= \int \calD (\bar\psi, \psi)\ (\bar\psi\psi)^{r} e^{h(\bar\psi\psi)}
	\left[1 + \bar\psi\psi \, g(\bar\psi\psi)\right]\, .
\end{multline}
\end{lemma}
By this lemma, $\calZ(t)$ can be written in terms of the sole mean-field 
variable $\bar\psi\psi$
\begin{equation}
  \label{eq:z-complete}
  \calZ(t) = \int \calD (\bar\psi, \psi)\,
  	\exp \left\{ t \,{\cal U} + n \frac {(\bar\psi\psi)^{k-1}} {(k-1)!} \right\}
	\left[ 1 - \frac {(\bar\psi\psi)^{k-1}}{(k-2)!} \right]
	.
\end{equation}
In order to perform an estimate for the asymptotic value of the integral for large $n$ we recall that for an analytic function $f$
\begin{equation}
	\int \calD (\bar\psi, \psi)\ f(\bar\psi\psi) \equiv n! \oint \frac {d\xi} {2 \pi i}\ \frac{f(\xi)}{\xi^{n+1}}
\end{equation}
where the integration contour in the complex plane is around the origin. We have the following complex integral representation form for the partition function $\calZ(t)$
\begin{multline}
  \label{eq:z-mf}
  \calZ(t) = n! \oint \frac {d\xi} {2 \pi i} \, \frac1{\xi^{n+1}}
  \exp \left\{ t \left[ \xi + (1 - k) \frac {\xi^k} {k!} \right] \right\}
	\\
	\exp \left\{ n \frac {\xi^{k-1}} {(k-1)!} \right\}
  \left[ 1 - \frac {\xi^{k-1}}{(k-2)!} \right]
	.
\end{multline}
Let us first work at fixed number of hypertrees,
in a micro--canonical ensemble in the physics terminology.

Expanding $\eqref{eq:z-mf}$ in powers of $t$ we obtain the number $\calZ_{p}$ of
spanning hyperforests on the complete $k$-uniform hypergraph which is the number of
states in the micro-canonical ensamble
\begin{multline}
  \label{eq:fp}
  {\cal Z}_p = \frac {n!} {p!} \oint \frac {d\xi} {2 \pi i} \, \frac1{\xi^{n+1}}
  \left[ \xi + (1 - k) \frac {\xi^k} {k!} \right]^p
  \\
  \exp\left\{ n \frac {\xi^{k-1}} {(k-1)!} \right\}
  \left[ 1 - \frac {\xi^{k-1}}{(k-2)!} \right]
  .
\end{multline}
Since we are interested in obtaining $ {\cal Z}_p$ in the thermodynamical limit $n \to \infty$
also for large values of $p$, we define $p = \alpha n$ with fixed $\alpha$
as $n, p \to \infty$. Changing the variable of integration to
$\eta = (k-1) \frac{\xi^{k-1}}{k!}$, we obtain the following 
integral expression:
\begin{equation}
  \label{eq:fp-AB}
  {\cal Z}_{\alpha n}  = \frac {n!} {\Gamma(\alpha n+1)}
  \left[ \frac{k -1}{k!} \right]^{n \frac {1-\alpha } {k - 1}}\, I(\alpha)
\end{equation}
where
\begin{equation}
I(\alpha) :=  \oint \frac{d\eta}{2 \pi i} \, A(\eta) \, e^{nB(\eta)} \label{intI}
\end{equation}
with
\begin{align}
	A(\eta) & = \frac{1 - k \eta}{\eta} \label{Axi} \\
	B(\eta) & = \frac k {k - 1} \eta + \alpha \log(1 - \eta) + \frac {\alpha - 1} {k - 1} \log \eta
	.
\end{align}
Please note that the factor $k - 1$ coming from the change of variable
in the integral is exactly compensated by the fact that a full turn
around the origin in the $\eta$ plane is equivalent to $k - 1$ turns of
the $\xi$ variable.

Precise estimates of integrals of this kind for $n \to \infty$ can be
obtained by the saddle point method (see \cite{FlaSed08} for a very complete
discussion of this method).

\subsection{The saddle point method} 
\label{sub:the_saddle_point_method}

A \emph{saddle point} of a function $B(\eta)$ is a point $\eta_0$ where $B'(\eta_0) =
0$, it is said to be a \emph{simple saddle point} if furthermore $B''(\eta_0)
\neq 0$. In this case it is easy to see that the equilevel lines divide a
neighborhood of $\eta_0$ in four regions where $\Re B(\eta)$ is alternately higher and
lower than the saddle point value $\Re B(\eta_0)$. We will refer to the two lower
regions as the \emph{valleys}.

Analogously, a \emph{multiple saddle point} has multiplicity $p$ if  all derivatives up to $B^{(p)}(\eta_0)$ are equal to zero while
$B^{(p+1)}(\eta_0) \neq 0$. In this case there are $p + 1$ higher and lower
regions.

When evaluating Cauchy contour integrals of the form~(\ref{intI}), saddle points of $B(\eta)$ play a central role in the asymptotic estimate for large $n$.  
The method essentially consists of two basic ingredients: an accurate choice
of the contour and Laplace's method for the evaluation of integrals depending
on a large parameter.

The contour has to be chosen to pass through a point which is a
global maximum of the integrand along the contour and that a neighborhood of
which (the \emph{central region}) dominates the rest of the contour (the
\emph{tails}) as $n$ grows. Since an analytic function cannot have an isolated
maximum, this implies that the contour should pass through a saddle point.

The existence of a contour surrounding the origin and that crosses a saddle
point along its direction of steepest descent requires that two of its valleys are
topologically connected and the region connecting them surrounds the
origin.

Once we have a contour, we proceed neglecting the tails and
 approximating the functions $A(\eta)$ and $B(\eta)$ with their Taylor series
about the chosen saddle point $\eta^*$. Then, after having absorbed the factor
$n$ into a rescaled variable $x = (\eta - \eta^*) / n^{1/(p+1)}$ (where $p$ is the multiplicity of the saddle
point), we can easily
obtain an asymptotic expansion of the integral in inverse powers of $n$.

We collect here the first few terms of the asymptotic 
expansion for the case of a simple saddle
\begin{equation}
  \label{eq:single-saddle}
I \simeq   \frac{e^{n B(\eta^{*})}}{\sqrt{2 \pi n B''(\eta^{*})}}
  \left[
		A(\eta^{*}) + \frac{1}{n} C(\eta^{*})
  	+ \frac{1}{n^2} D(\eta^{*})
		+ O\left(\frac{1}{n^3}\right)
  \right]
\end{equation}
where the terms in the square brackets with half-integer inverse-power of $n$ vanish, and of a double saddle
\begin{equation}
	\label{eq:double-saddle}
I \simeq   \frac{e^{n B(\eta^{*})}}{n^{\frac{1}{3}} B^{(3)}(\eta^{*})^{\frac{1}{3}}}
	\left[
 		\gamma_{0} \, A(\eta^{*}) + \frac{1}{n^{\frac{1}{3}}} \tilde C(\eta^{*})
  	+ \frac{1}{n} \tilde D(\eta^{*})
		+ O\left(\frac{1}{n^{\frac{4}{3}}}\right)
  \right]
\end{equation}
where the terms in the square brackets with powers $n^{-(l+\frac{2}{3})}$, with integer $l$,  vanish. In these formulae $C$, $\tilde C$, $D$, and $\tilde D$ are rational
functions of $A(\eta^*)$, $B(\eta^*)$ and their derivatives, whose expression is reported in the Appendix~\ref{sec:appendix}, together with the value of the constant~$\gamma_0$.

For our integral~(\ref{intI}) in the large $n$ limit the relevant saddle-point equation
$B'(\eta)= 0$ has two solutions, $\eta_a$ and $\eta_b$:
\begin{align}
	\label{eq:saddles}
	\eta_a = \frac{1}{k} \qquad \eta_b = 1 - \alpha
	.
\end{align}
If $\alpha \neq \alpha_c \equiv (k - 1)/k$ the two solutions are distinct and
correspond to simple saddle points. To understand which one is relevant to our
discussion we need to study the landscape of the function $B(\eta)$ beyond the
neighborhood of the saddles.

\begin{figure}[tbp
]
	\centering
\setlength{\unitlength}{50pt}
\begin{picture}(6.2,4.6)
\put(0,0){\includegraphics[scale=1]{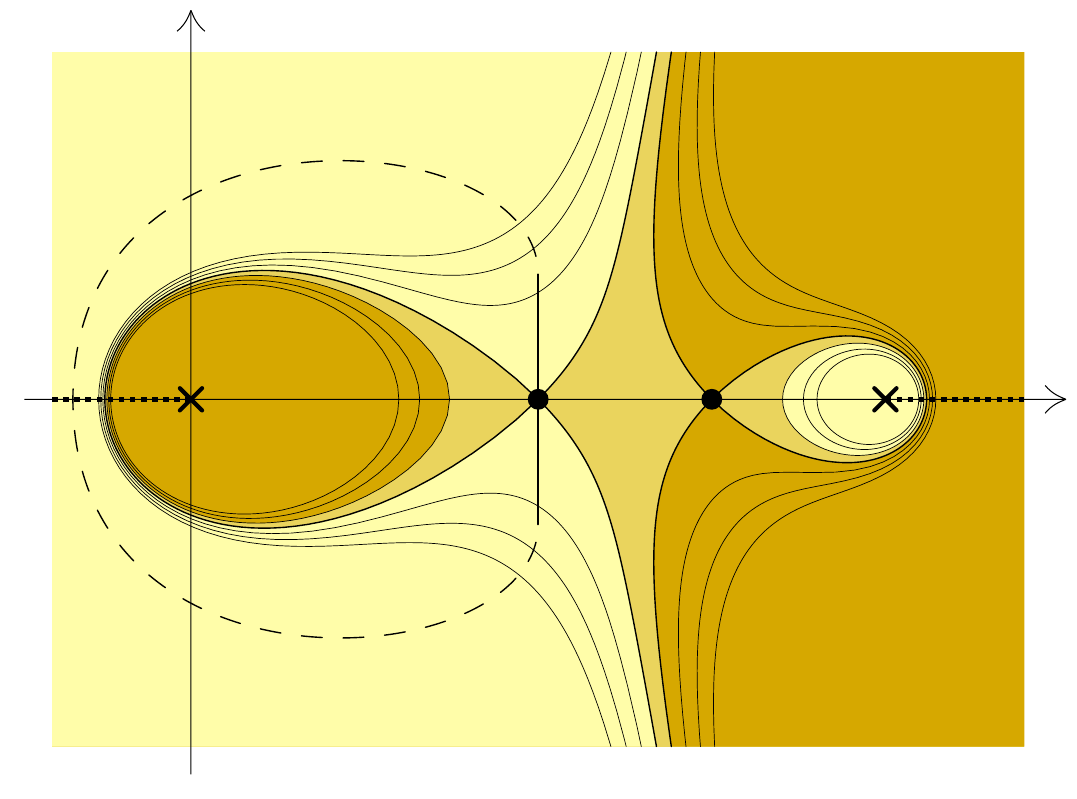}}
\put(2.89,2){$\eta_a$}
\put(4.03,2){$\eta_b$}
\put(0.9,2){$0$}
\put(4.95,2){$1$}
\end{picture}

\caption{\small Contour levels for $\Re B(\eta)$ when $\alpha < \alpha_c$. More
  precisely, the figure shows the case $k=2$ and $\alpha = \frac{1}{2} \alpha_c$.
The two bold contour lines describe the level lines of $\Re B(\eta)$ for
the values at the two saddle points (located at the bullets). Darker
tones denote higher values of $\Re B(\eta)$. The crosses and the dotted lines
describe the cut discontinuities due to logarithms in $B(\eta)$. The
dashed path surrounding the origin going through one of the saddle
points is an example of valid integration contour, and the solid
straight portion of the path describes an interval in which the
perturbative approach is valid.
\label{fig:SaddlesLow}
}
\end{figure}
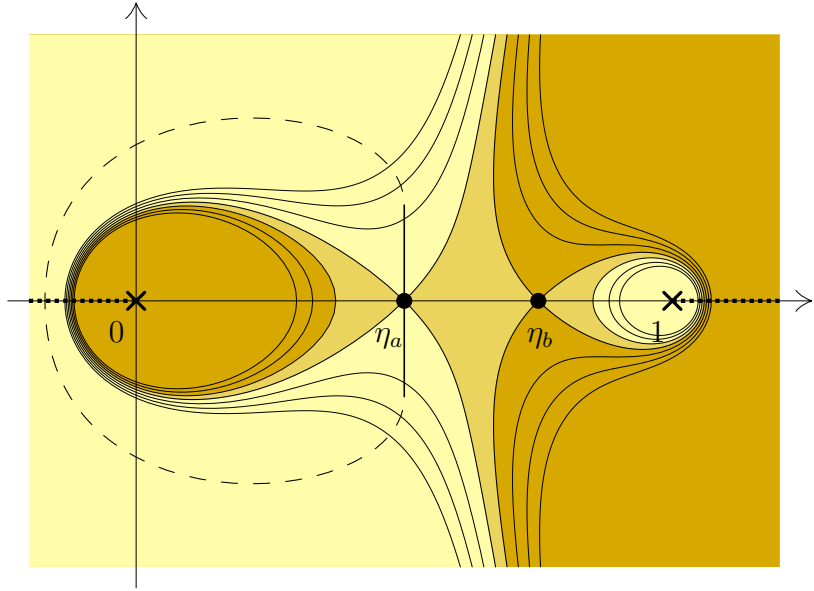

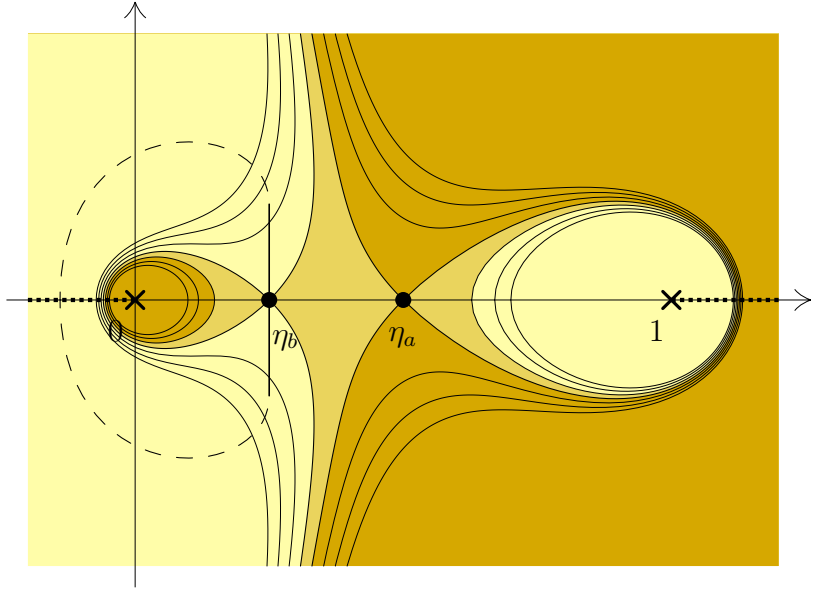
\begin{figure}[htbp]
	\centering
\setlength{\unitlength}{50pt}
\begin{picture}(6.2,4.6)
\put(0,0){\includegraphics[scale=1]{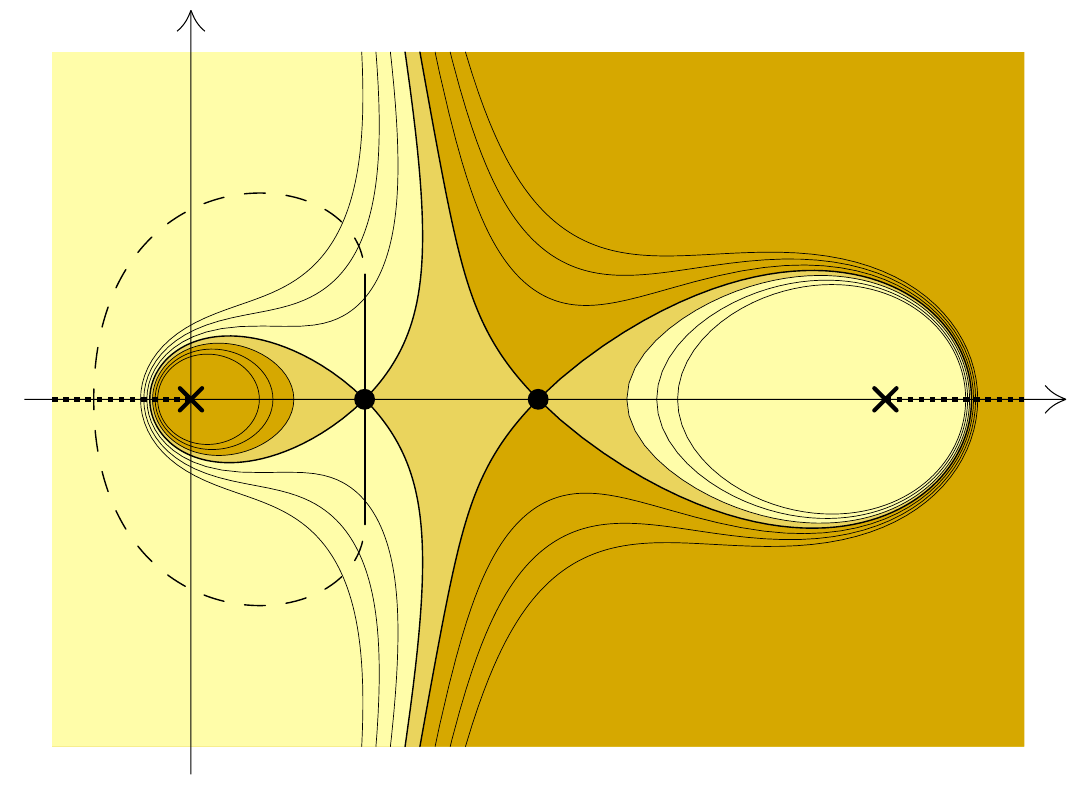}}
\put(3,2){$\eta_a$}
\put(2.13,1.98){$\eta_b$}
\put(0.9,2){$0$}
\put(4.95,2){$1$}
\end{picture}

\caption{\small Contour levels for $\Re B(\eta)$ when $\alpha > \alpha_c$. More
  precisely, the figure shows the case $k=2$ and $\alpha = \frac{3}{2} \alpha_c$.
Description of notations is as in figure~\ref{fig:SaddlesLow}.
\label{fig:SaddlesHigh}
}
\end{figure}

\begin{figure}[htbp]
	\centering
\setlength{\unitlength}{50pt}
\begin{picture}(6.2,4.6)
\put(0,0){\includegraphics[scale=1]{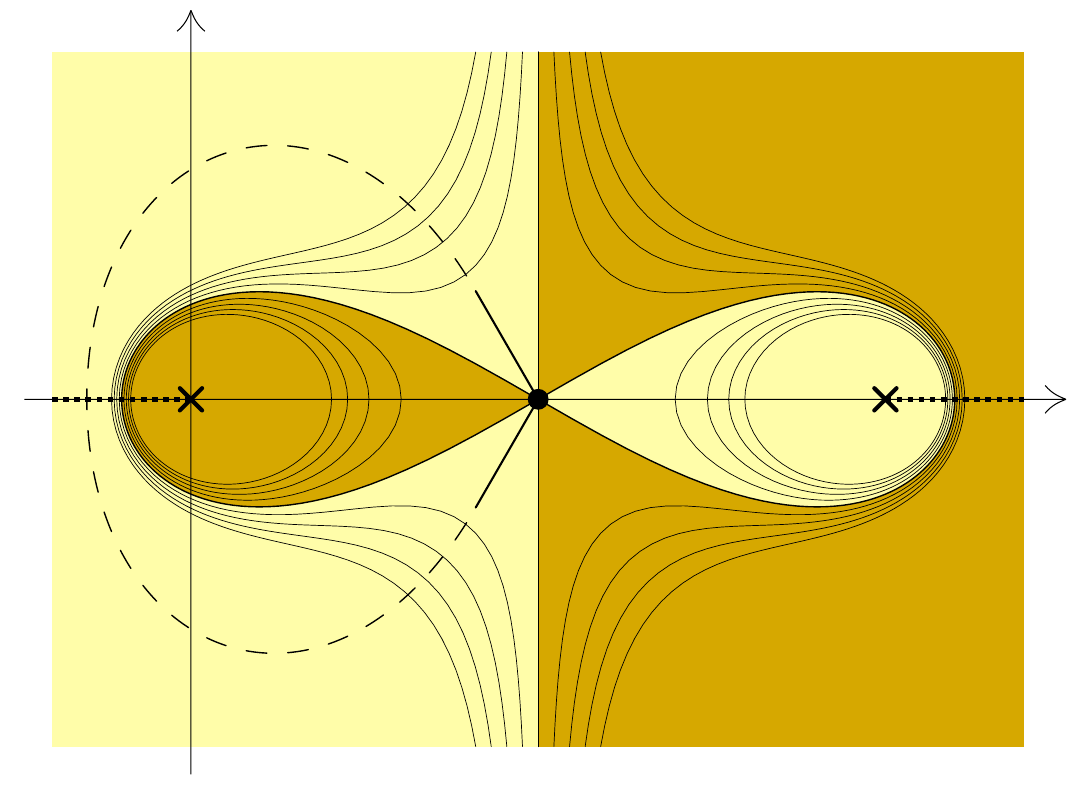}}
\put(3.165,2){$\eta_a \equiv \eta_b$}
\put(0.9,2){$0$}
\put(4.95,2){$1$}
\end{picture}

\caption{\small Contour levels for $\Re B(\eta)$ when $\alpha = \alpha_c$. More
  precisely, the figure shows the case $k=2$.
Description of notations is as in figure~\ref{fig:SaddlesLow}.
\label{fig:SaddlesCritical}
}
\end{figure}

In our specific case, as illustrated in figures from~\ref{fig:SaddlesLow} to
\ref{fig:SaddlesCritical}, when $\alpha < \alpha_c$ among the two saddles only
$\eta_a$ is accessible, while,  if $\alpha > \alpha_c$, only
$\eta_b$ is so. When $\alpha = \alpha_c$ the two saddle points coalesce into a
double saddle  
point, thus with three valleys, having steepest-descent directions
$e^{\frac{2 \pi i k}{3}}$, with $k=0$, 1, 2. Of these valleys, the
ones with the appropriate global topology are those with indices $k=1$
and $2$.

As a first result of this discussion, in order to study the asymptotic behaviour of
$\calZ_{\alpha n}$ will need to distinguish two different phases, and a critical point, upon the value
of $\alpha$ being below, above or equal to $\alpha_c$.

We will name the phases with a smaller and a larger number of  hypertrees, respectively, the {\em low temperature} and {\em high temperature} phase,
the reason being that, as we shall see, in the low temperature phase there is a spontaneous symmetry breaking and the appearance of a non-zero residual {\em magnetization}.


\subsubsection{Low temperature phase}
\label{sec:case-a}

In the case $\alpha < \alpha_c$  the relevant saddle point is $\eta_a = 1/k$. See Fig.~\ref{fig:SaddlesLow}. Since $A(\eta_a) = 0$, we are in the  case in which the
leading order of \eqref{eq:single-saddle} vanishes and the next order has to
be considered. The expansion of $A(\eta)$ and $B(\eta)$ in a neighborhood of the
saddle $\eta_{a}$ is as follows:
\begin{align}
  \label{eq:2}
  A(\eta_a + u) & \simeq - k^2 u + k^3 u^2 + O(u^3) \\
  \label{eq:4}
 \begin{split}
  B(\eta_a + u) & \simeq 
  \frac{1+(1-\alpha) \log k}{k \, \alpha_c}  + \alpha \log \alpha_c 
  + k\, \frac{\alpha_c - \alpha}{\alpha_{c}^2} \frac{u^2}{2}
  \\ 
  & \quad 
  - \left[ k^2 \frac {1-\alpha} {\alpha_c} + \frac{\alpha}{\alpha_c^3}\right]
  \frac {u^{3}}{3} + O(u^{4})\, .
  \end{split}
\end{align}
Using formula \eqref{eq:single-saddle} we obtain for \eqref{eq:fp} the following asymptotic expression:
\begin{align}
  \label{eq:18}
  {\cal Z}_{\alpha n} \simeq &
	\frac {n!} {\Gamma(\alpha n+1)}
        \frac {\alpha \sqrt{k-1}} {\sqrt{2 \pi n^3 }}
	\frac {e^{\frac{n}{ k - 1}}  \, \left(\frac{k-1}{k}\right)^{ n \alpha -1}}
	{ [(k-2)!]^{n \frac {1 - \alpha}{k - 1}}}       
	\left(1 - \frac{k \alpha}{k-1}\right)^{-5/2}
	\\
  \label{eq:27}
  \simeq &
	\frac {n^{n-2}} {(\alpha n)^{\alpha n - \frac{1}{2}}}
        \sqrt{\frac {k-1} {2 \pi }}
	\frac {e^{\left( \alpha - \frac{k-2}{k-1}\right) n} \, \left(\frac{k-1}{k}\right)^{ n \alpha -1}} 	{ [(k-2)!]^{n \frac {1 - \alpha}{k - 1}}}       
          \left(1 - \frac{k \alpha}{k-1}\right)^{-5/2}
\end{align}
where in the second line we used the Stirling formula to approximate the large factorial $n!$.
In a previous work~\cite{BedCarSpo08} we already gave an asymptotic formula
for the number of forests with a given number $p$ of connected components.
That formula has been obtained keeping $p$ fixed while doing the limit $n \to
\infty$, in the notation of this paper this means taking $\alpha$
infinitesimal. By setting $ \alpha n \to p$ in \eqref{eq:18} and using
\begin{equation}
\frac{\alpha}{ \Gamma(\alpha \, n +1)} = \frac{\alpha }{\alpha n \,\Gamma(\alpha n)} = \frac{n^{-1}}{(p-1)!}
\end{equation}
 and then taking
the limit $\alpha \to 0$, we can re-obtain the result in~\cite{BedCarSpo08} by using again the Stirling formula to approximate the large factorial $n!$:
\begin{equation}
  \label{eq:28}
  {\cal Z}_p \simeq \frac{n^{n - 2}}{e^{n \frac {k - 2} {k - 1}}} \frac{
    \sqrt {k - 1}} {\left[ (k-2)! \right]^{\frac{n-p}{k-1}}}
  \frac1{(p-1)!} \left( \frac{k-1}k \right)^{p-1}\, .
\end{equation}

\subsubsection{High temperature phase}
\label{sec:case-b}

When $\alpha_c< \alpha < 1$ the relevant saddle point changes into $\eta_b = 1
- \alpha$ (see Fig.~\ref{fig:SaddlesHigh}) where the functions $A(\eta_b+u)$ and $B(\eta_b+u)$ can be approximated at $O(u^4)$
with
\begin{align}
  A(\eta_b + u) & \simeq k\,\frac{\alpha - \alpha_{c}}{1 - \alpha}
  - \frac{u}{(\alpha - 1)^2} - \frac{u^2}{(\alpha - 1)^3}
  - \frac{u^3}{(\alpha - 1)^4}
  \\
  \begin{split} 
  B(\eta_b + u) & \simeq 
  \frac{1 - \alpha}{\alpha_c} \left[ 1 - \frac{1}{k}\log(1 - \alpha)\right] + \alpha \log{\alpha}
 \\ & \;
 + \frac{1}{\alpha \, \alpha_{c}} \frac{\alpha - \alpha_c}{1 - \alpha} \frac{u^2}{2}
 - \left[ \frac{1}{\alpha^2} + \frac{1}{k\, \alpha_c\,(1-\alpha)^2}\right]
  \frac{u^3}{3} 
  \end{split}
\end{align}
The situation is quite analogous to the previous one, with the exception that  $A(\eta_b) \neq 0$, and using formula
\eqref{eq:single-saddle} we obtain for \eqref{eq:fp} the following asymptotic 
expression:
\begin{align}
  \label{eq:7}
  {\cal Z}_{\alpha n} \simeq &
	\frac {n!\, \alpha^{\alpha n}\,(k-1)} {\Gamma(\alpha n+1)}
	\left[\frac {e^{k} }{ k\,(1-\alpha)\,
		 (k-2)! }\right]^{n \frac {1 - \alpha} {k - 1}} \sqrt{\frac {\alpha}{2 \pi n (1 - \alpha)}}\nonumber\\
	& \qquad 
	\left( \frac{\alpha k}{k-1} -1\right)^{1/2}
	\\
	 \simeq &
	\frac {n^{(1-\alpha)n}} {\sqrt{2 \pi n \frac{1-\alpha}{k-1}}}
        \left[\frac {e}{ k\,(1-\alpha)\,
		 (k-2)! }\right]^{n \frac {1 - \alpha} {k - 1}}	\,	\left( \alpha k - k + 1\right)^{1/2}\, .
\end{align}
Remark that the saddle point method cannot be applied when $\alpha=1$, but if we  replace
\begin{equation}
\frac{e^{n \frac {1 - \alpha} {k - 1}}}{\sqrt{2 \pi n \frac{1-\alpha}{k-1}}} \simeq \frac{n^{n \frac {1 - \alpha} {k - 1}}}{\Gamma\left( n \frac {1 - \alpha} {k - 1} + 1 \right)}
\end{equation}
for $\alpha \simeq 1$ we get
\begin{equation}
{\cal Z}_{ n} \simeq 1 \, .
\end{equation}
as we should.

\subsubsection{The critical phase}
\label{sec:critical-phase}

When $\alpha$ is exactly $\alpha_c = (k - 1)/k$, the saddle points $\eta_b$ and
$\eta_a$ coalesce into a double saddle point in $\eta_a$ in which
the second derivative vanishes along with the first one. The expansion of
$A(\eta)$ and $B(\eta)$ are as in (\ref{eq:2})-(\ref{eq:4}) with $\alpha =
\alpha_{c}$:
\begin{align}
  A(\eta_c + u) & \simeq - k^2 u + k^3 u^2 + O(u^3) \\
  \begin{split}
  B(\eta_c + u) & \simeq 
  \frac{1}{k-1} + \frac{1}{k(k-1)} \log (k-1) + \frac{k-2}{k-1} \log k
  \\ & \quad 
  - \frac{k^3}{(k - 1)^2} \frac{u^{3}}{3} + \frac{k^4 (k - 2)}{(k - 1)^3} \frac{u^{4}}{4} + O(u^{5})
  \end{split}
\end{align}
Using \eqref{eq:double-saddle} we obtain the following result
\begin{align}
  \label{eq:10}
  {\cal Z}_{\alpha_c n} = & \frac {n!} {\Gamma\left( \frac{k-1}{k} n +1 \right)} \frac{e^{\frac {n} {k - 1} }\left(\frac{k-1}{k}\right)^{\frac{k-1}{k} n}} { \left[ (k - 2)!
    \right]^{\frac n {k(k - 1)}}} \, \frac {3^{1/6} \Gamma (2/3) (k -
    1)^{4/3}} {2 \pi \, n^{2/3}}\\
    \simeq & \,n^{\frac{n}{k}}\,  \left[\frac{e}{(k-2)!}\right]^{\frac{n}{k(k-1)}}\, \frac {3^{1/6} \Gamma (2/3) (k -
    1)^{4/3}} {2 \pi \, n^{2/3}}\, .
\end{align}
This formula, for $k=2$,  can be, in principle, compared with the result presented in~\cite[Proposition VIII.11]{FlaSed08}, but unfortunately there is a discrepancy in the numerical pre-factor.

%

\subsection{The canonical ensemble}

According to the definition~(\ref{eq:fp-AB}) for  the number of forests with $p = \alpha n$ trees $\mathcal Z_{\alpha n}$, by evaluating the integral $I$ defined in~(\ref{intI}) by the saddle-point method, when the saddle point $\eta^*(\alpha)$ is simple and thus away from the critical point $\alpha_c$,  we get  the following asymptotic expansion for large number of vertices $n$:
\begin{equation}
	\label{eq:1}
	\mathcal Z_{\alpha n} 
	\simeq \frac{n! }{\Gamma(\alpha n+1)} \,  \left[ \frac{k-1}{k!} 
	\right]^{n \frac{1-\alpha }{k - 1}} \frac{e^{n B(\eta^{*})}}{\sqrt{2 \pi n B''(\eta^{*})}}
  \left[ A(\eta^{*}) + \frac{1}{n} C(\eta^{*}) \right]\, .
\end{equation}
We define the entropy density $s(\alpha)$ as
\begin{equation}
	s(\alpha) = \frac{1}{n} \log \frac{\mathcal Z_{\alpha n}}{n!}
\end{equation}
so that we can recover the partition function $\mathcal Z(t)$ by a Legendre transformation
\begin{align}
	\mathcal Z(t) & = \sum_p \mathcal Z_p\, t^p \\
	& \simeq  \int_{0}^{1} d\alpha\, \mathcal  Z_{\alpha n}\, t^{n\alpha} \\
	& = n! \int_{0}^{1} d\alpha\, \exp
	\left\{ n [s(\alpha) + \alpha \log t] \right\}
\end{align}
that can be evaluated for large $n$ once more by the saddle-point method. Calling 
$\bar\alpha(t)$ the mean number of trees at given $t$, we have:
\begin{equation}
	s'(\bar\alpha(t)) + \log t = 0\, .
\end{equation}
From \eqref{eq:1} we see that $s(\alpha)$ still has an $\alpha$-dependent leading
order in $n$
\begin{equation}
	s(\alpha) \simeq - \alpha \log n - \alpha \log \alpha + \alpha +
	\frac{\alpha - 1}{k - 1} \log \left[ \frac{k!}{k-1} 
	\right] + B(\eta^*(\alpha))
\end{equation}
that would shift the solution down to 0. By the rescaling  
\begin{equation}
t = n\, \tilde t
\end{equation}
which is usual in the complete graph, in order to obtain a correct thermodynamic scaling, we can reabsorb this factor. The saddle-point equation now reads
\begin{equation}
	s'(\bar\alpha) + \log n + \log \tilde t = 0
\end{equation}
whose solution is
\begin{equation}
	\bar\alpha = \begin{cases}
		\frac{k-1}{k}\, \left[(k-2)!\right]^\frac{1}{k-1} \, \tilde t
		& \hbox{for\quad}\tilde t < \tilde t_c \\
		1 - \frac{k-1}{k!} \frac{1}{\tilde t^{k-1}} & \hbox{for\quad} \tilde t > \tilde t_c
	\end{cases}
\end{equation}
where $\tilde t_c = [(k-2)!]^{-1/(k-1)}$. And by inversion
\begin{equation}
	\tilde t = \begin{cases}
		\frac{k}{k-1}\, \left[(k-2)!\right]^{-\frac{1}{k-1}} \, \bar\alpha & \hbox{for\quad} \bar\alpha < \bar\alpha_c \\
	\left( \frac{k - 1}{k!}\frac{1}{ 1 - \bar\alpha}\right)^\frac{1}{k-1} & \hbox{for\quad} \bar\alpha > \bar\alpha_c\, .
	\end{cases}
\end{equation}
In the ordinary graph case, this means
\begin{equation}
	\bar\alpha = \begin{cases}
		\frac{\tilde t}{2} & \tilde t < 1 \\
		1 - \frac{1}{2\tilde t} & \tilde t > 1
	\end{cases}
	\qquad
	\tilde t = \begin{cases}
		2 \bar\alpha & \bar\alpha < \frac{1}{2} \\
		\frac{1}{2(1 - \bar\alpha)} & \bar\alpha > \frac{1}{2}\, .
	\end{cases}
\end{equation}


\section{Size of the hypertrees}
\label{sec:t}

We have shown in the previous sections that the system admits two
different phases. We want now to characterize these two regimes.  Our
field-theoretical approach provides us a full algebra of observables,
as polynomials in the Grassmann fields, which we could study
systematically. However, it is interesting to note that some of these
observables have a rephrasing in terms of combinatorial properties of
the forests (cfr.~\cite{CarSokSpo07}). Furthermore, we are induced by the results of~\cite{DenGarSok07}
 to investigate the possibility of a transition of percolative
nature, with the emergence of a giant component in the {\em typical}
forest for a given ensemble.

A possibility of this sort is captured by the mean square size of the
trees in the forest, as the following argument shows at least at a
heuristic level. If we have all trees with size of order 1 in the
large $n$ limit, (say, with average $a$ and variance $\sigma$ both of order 1), then the sum of the squares
of the sizes of the trees in a forest scales as $(a+\sigma^2/a) n$. If, conversely,
in the large $n$ limit one tree occupies a finite fraction $p$ of the
whole graph, the same sum as above would scale as $p^2 n^2 +
\mathcal{O}(n)$.

Furthermore, it turns out that the combinatorial observable above has
a very simple formulation in the field theory, corresponding to the
natural susceptibility for the fermionic fields, as we will see in a
moment.

Let's start our analysis with the un-normalized expectation
\begin{align}
t\,\langle \bar\psi_i \psi_i \rangle =  & \,t\,\int \calD (\bar\psi, \psi)\,\bar\psi_i \psi_i\, \exp \left( - {\cal H} \right) \\
=  &  \, {\cal Z}(t)
\end{align}
because the insertion of the operator $ \bar\psi_i \psi_i$ simply marks the vertex $i$ as a root of a hypertree, and in a spanning forest every vertex can be chosen as the root of a hypertree. 
If we now sum over the index $i$ we gain a factor $|T|$ for each hypertree. Therefore we have:
\begin{equation}
	\label{eq:t-lin}
t \, \<\bar\psi\psi \>  = t \,\sum_{i \in V} \<\bar\psi_i\psi_i \>  =
  \sum_{F \in \cal{F}} t^{K(F)} \sum_{T \in F} |T| \prod_{A \in T} w_A   = n\, {\cal Z}(t) 
\end{equation}
as in each spanning hyperforest the total size of the hypertrees is the number of vertices in the graph, that is $n$.
By expanding in the parameter $t$ and by taking the $p$-th coefficient we get the relation
\begin{equation}
\frac{1}{{\cal Z}_p} \frac{\langle \bar\psi \psi \, {\cal U}^{p-1}\rangle_{t=0}}{(p-1)!}  = n\, . \label{W1}
\end{equation}

For the  un-normalized two-point function
\begin{equation}
\langle \bar\psi_i \psi_j \rangle =  \int \calD (\bar\psi, \psi)\,\bar\psi_i \psi_j\, \exp \left( - {\cal H} \right)\, 
\end{equation}
we know (see~\cite{CarSokSpo07}) that
\begin{equation}
t\,  \< \bar\psi_i \psi_j \>  = \sum_{\substack{F\in \calF \\ i,j \text{ connected}}}
  t^{K(F)} \sum_{T \in F} \prod_{A \in T} w_A\, .
\end{equation}
As $i$ and $j$ are connected if they belong to the same hypertree, 
if we sum on both indices $i$ and $j$ we gain a factor $|T|^2$ for each hypertree
\begin{equation}
	\label{eq:t-square}
t\,\<(\bar\psi, \J \psi) \> =  t\,\sum_{i,j \in V} \<\bar\psi_i\psi_j \>  =
  \sum_{F\in \calF} t^{K(F)} \sum_{T \in F} |T|^2 \prod_{A \in T} w_A 
\, .
\end{equation}
The effect of this observable is to introduce an extra weight for hypertrees in the spanning forests which is
the square of its size.

The average of the square-size of hypertrees  in the microcanonical ensemble of hyperforests with fixed number $p$ of hypertrees  is easily obtained from the previous relation by expanding in the parameter $t$ and by taking the $p$-th coefficient, so that
\begin{equation}
\<\, |T|^2\>_p :=  \frac{1}{\calZ_{p}} \frac{\< (\bar\psi, \J \psi)\, \calU^{p-1} \>_{t=0}}
  {(p - 1)!}\, .
\end{equation}

The very same method of the preceding Section can be used to evaluate
this quantity. Still in a mean-field description, we have:
\begin{align}
  \label{eq:47}
  & \< (\bar\psi, \J \psi) \, \calU^{p-1} \>_{t=0} = \\
  & = 
  \int \calD (\bar\psi, \psi)\ (\bar\psi, \J \psi) \ \calU^{p-1} 
  \exp\left\{ n \frac {(\bar\psi\psi)^{k-1}} {(k-1)!} \right\}
  \left[ 1 - (\bar\psi, \J \psi) \frac {(\bar\psi\psi)^{k-2}}{(k-2)!} \right]
  \\
  & = \int \calD (\bar\psi, \psi)\ \bar\psi\psi \ \calU^{p-1}
  \exp\left\{n \frac {(\bar\psi\psi)^{k-1}} {(k-1)!} \right\}
  \\
  & = n! \oint \frac{d\xi}{2\pi i} \,
    \frac{1}{\xi^{n + 1}} \ \xi
    \left[ \xi + (1 - k) \frac {\xi^k} {k!} \right]^{p-1}
    \exp \left\{n \frac {\xi^{k-1}} {(k-1)!} \right\} \\
  & = n!
  \left[ \frac{k-1}{ k!}\right]^{n \frac {1-\alpha } {k - 1}}
  \oint \frac{d\eta}{2 \pi i}\ \tilde{A}(\eta)
  \ e^{n B(\eta)}
  ,
\end{align}
where now
\begin{equation}
  \tilde{A}(\eta)  = \frac{1}{\eta (1 - \eta)}
  ,
\end{equation}
and $B(\eta)$ is the same as before. To evaluate this integral we again use the
saddle point method. Please note that since the function $B(\eta)$ is
unchanged so are the saddle points.

Using the general expansion for $p= \alpha n$~\eqref{eq:single-saddle} we have
\begin{align}
	\label{eq:t-square-saddle}
	\<\, |T|^2\>_{\alpha n}  = \frac{1}{\calZ_{\alpha n} }\,\frac{\< (\bar\psi, \J \psi)\, \calU^{\alpha n-1}\>_{t=0}}{\Gamma(\alpha n)} =
	\alpha n\, \frac{
	\tilde{A}(\eta^*) + \frac1n \tilde{C}(\eta^*) + O\left(\frac{1}{n^2}\right)
	}{
	A(\eta^*) + \frac1n C(\eta^*) + O\left(\frac{1}{n^2}\right)}
	.
\end{align}
Now in the low temperature phase we have $A(\eta_{a}) = 0$ so in order to get the leading term we need $C(\eta^*)$ and as
\begin{equation}
	\tilde{A}(\eta_{a}) = \frac{k^2}{k-1} \quad \text{and} \quad C(\eta_{a}) = \frac{\alpha (k - 1)}{(\alpha - \alpha_c)^2},
\end{equation}
\eqref{eq:t-square-saddle} at leading order gives
\begin{equation}
\<\, |T|^2\>_{\alpha n} \, \simeq \, \alpha n^2 \,\frac{\tilde{A}(\eta_{a})}{C(\eta_{a})}
	= n^2 \left( \frac{\alpha_c - \alpha}{\alpha_c} \right)^2 \label{giant}
\end{equation}
so that, as soon as $\alpha < \alpha_c$, a giant hypertree appears in
the typical forest, which occupies on average a fraction
$1-\alpha/ \alpha_c$ of the whole graph. In the high temperature instead we have
\begin{equation}
	\tilde{A}(\eta_b) = \frac{1}{\alpha(1 - \alpha)}
	\quad \text{and} \quad
	A(\eta_{b}) = k \frac{\alpha - \alpha_{c}}{1 - \alpha}
	,
\end{equation}
giving (always at leading order)
\begin{equation}
\<\, |T|^2\>_{\alpha n}\, \simeq	 \, \alpha n \, \frac{\tilde{A}(\eta_{b})}{A(\eta_{b})}
	= \frac{n}{k } \frac{1}{\alpha - \alpha_{c}}\, .
\end{equation}
So that
\begin{equation}
\lim_{n\to \infty} \frac{1}{n^2}\, \<\, |T|^2\>_{\alpha n}  = 
\begin{cases}
 \left( \frac{\alpha_c - \alpha}{\alpha_c} \right)^2 & \hbox{for\, } \alpha \leq \alpha_c \\
 0 & \hbox{for\, } \alpha \geq \alpha_c 
\end{cases}
\end{equation}
is an order parameter, but it is represented as the expectation value of a non-local operator. 
We shall see in the next Section how to construct a local order parameter.


\section{The symmetry breaking} 
\label{sec:the_symmetry_breaking}

In this section we will describe the phase transition in terms of the breaking of the
global $\osp(1|2)$ supersymmetry. According to the general strategy (see for example~\cite{Zinn-Justin})  let's add an exponential weight with an external source
$h$ coupled to the variation of the fields \eqref{eq:osp-fermionica}  and \eqref{eq:osp-fermionicb}:
\begin{equation}
	h\,  \sum_{i \in V} (1 - t \,\bar\psi_{i} \psi_{i}) \, = \, h\, (n - t\, \bar\psi\psi)
	,
\end{equation}
The partition
function becomes now:
\begin{equation}
  \label{eq:45}
  \calZ (t, h)  = \int \calD (\bar\psi, \psi)\ 
	e^{- \calH[\psi,\bar\psi] - h(n - t \bar\psi\psi)} 
  .
\end{equation}
We have chosen to add the exponential weight with a minus sign because in this way when  $t$ is sent to zero with the product $h\,t$ kept
fixed, we get, aside from a vanishing trivial factor, the generating function of rooted hyperforests.

More generally, for finite $t$ and $h$, we have that $\calZ(t,h)$ can be expressed as a sum over spanning hyperforests with a modified weight
\begin{equation}
\calZ(t,h)  =  \sum_{F \in {\cal F}} \prod_{T\in F} t\, e^{- h |T|}\,  (1+ h\,   |T|)
\end{equation}
which is always positive, at any $n$,  only for $h\ge 0$.

On the $k$-uniform complete hypergraph the partition function~(\ref{eq:45}) is expressed
\begin{multline}
  \calZ (t, h) = \int \calD (\bar\psi, \psi)\ \exp
  \left\{ t \, {\cal U}
  + h t \, \bar\psi\psi
  \right\}
  \\
  \exp \left\{ - n h + n \frac{(\bar\psi\psi)^{k-1}}{(k-1)!} \right\}
  \left[ 1 - \frac{(\bar\psi\psi)^{k-1}}{(k-2)!} \right]
\end{multline}
To work in the micro-canonical ensemble we again expand in powers of $t$
\begin{equation}
	\calZ(t, h) = \sum_{p = 0}^{n} \calZ_p(h) \, t^p,
\end{equation}
where each term of the above expansion gives the partition function at fixed 
number of components:
\begin{align}
	\calZ_p(h) = & \frac{1}{p!}\,\int \calD (\bar\psi, \psi)\
	\left(\,{\cal U} + h \bar\psi\psi 
	\right)^{p}
	\nonumber \\
	& \quad \exp\left\{- n h + n \frac{(\bar\psi\psi)^{k-1}}{(k-1)!}\right\}	
	\left[ 1 - \frac{(\bar\psi\psi)^{k-1}}{(k-2)!} \right]
	 \\
       = & \frac{e^{-nh}}{p!}\, \langle \left(\,{\cal U} + h \bar\psi\psi 
	\right)^{p} \rangle_{t=0}\, .
\end{align}
Following the very same steps of the previous section, we can write this
expression in terms of a complex integral:
\begin{equation}	
	\label{eq:zp-with-h}
	\calZ_{p}(h) = \frac {n!} {\Gamma(\alpha n+1)}
	\left[\frac{ k-1}{ k!}\right]^{n \frac {1-\alpha} {k - 1}}\, I(\alpha,h)
\end{equation}
with
\begin{equation}
I(\alpha,h) :=   \oint \frac {d\eta} {2 \pi i} \, A(\eta) \, e^{n B(\eta, h)}
\end{equation}
where $A(\eta)$ is the same as in~(\ref{Axi}) while
\begin{equation}
  \label{eq:54}
  B(\eta, h)  :=  -  h + \frac k {k - 1} \eta + \alpha \log(1 + h - \eta)
  + \frac {\alpha - 1} {k - 1} \log \eta
  .
\end{equation}
$I(\alpha, h)$ can be again evaluated with the same technique as above.
Let's call $\eta^{*}(h)$ the position of the relevant saddle point, which is
the accessible solution of the saddle point equation
\begin{equation}
	\left. \frac{\de }{ \de \eta}  B(\eta, h)\,\right|_{\eta =\eta^*(h)}  = 0
	\, .
\end{equation}
If $h > 0$ the two solutions are real valued and distinct for every value of $\alpha$ and the
accessible saddle is simple and turns out to be always the one closer to the origin.

In the following we are going to consider all the functions $A$, $B$, $C$ and
$D$ as evaluated on $\eta^{*}(h)$ and therefore as functions of the single
parameter $h$.
\begin{align}
	& A(h) \equiv A(\eta^{*}(h)) \quad & B(h) \equiv B(\eta^{*}(h), h) \\
	& C(h) \equiv C(\eta^{*}(h), h) \quad & D(h) \equiv D(\eta^{*}(h), h)
\end{align}
The  asymptotic behaviour of \eqref{eq:zp-with-h} is given by the general 
formula \eqref{eq:single-saddle}:
\begin{equation}
  \calZ_{\alpha n}(h) \propto \frac{e^{n B(h)}}{\sqrt{2 \pi n B''(h)}}
  \left[
		A(h) + \frac{C(h)}{n} + O\left(\frac{1}{n^2}\right) 
	\right]\, .
\end{equation}
The density of entropy is obtained by taking the logarithm of the partition function
$\calZ_{\alpha n}(h)$
\begin{equation}
	s(\alpha, h) :=  \frac{1}{n} \log \frac{\calZ_{\alpha n}(h)}{n!}\, .
\end{equation}
The magnetization is then the first derivative of the entropy
\begin{align}
m(\alpha,h) = - \frac{\de s  }{ \de h} = & 1 -  \alpha \, \frac{ \langle \bar\psi \psi \,\left( {\cal U} + h \bar\psi \psi\right)^{\alpha n -1}\rangle_{t=0}}{\langle \left( {\cal U} + h \bar\psi \psi\right)^{\alpha n}\rangle_{t=0}} \\
= & 1 -  \alpha \,n\,  \frac{ \langle \bar\psi_i \psi_i \,\left( {\cal U} + h \bar\psi \psi\right)^{\alpha n -1}\rangle_{t=0}}{\langle \left( {\cal U} + h \bar\psi \psi\right)^{\alpha n}\rangle_{t=0}}
\end{align}
which is written as the expectation of a local operator, and if we set $h=0$ in this formula we get
\begin{equation}
 m(0) = 1 - \frac{1}{{\cal Z}_{\alpha n}(0)} \,\frac{ \langle \bar\psi_i\psi_i \, {\cal U}^{\alpha n-1}\rangle_{t=0}}{\Gamma(\alpha n)} = 0
\end{equation}
because of~(\ref{W1}).
In order to evaluate first the limit of large number of vertices we use the asymptotic expression for  $\calZ_{\alpha n}(h)$ to get
\begin{multline}
	m(\alpha,h) = 
	-\frac{\de B(h) }{ \de h}
	+\frac{1}{2n} \frac{1}{B''(h)} \frac{\de B''(h) }{ \de h}
	\\
	- \frac{1}{n} 
	\frac{\frac{\de A(h) }{\de h}
	+ \frac1n \frac{\de C(h) }{ \de h}
	+ O(\frac1{n^2})}
	{ A(h) + \frac1n C(h) + O\left(\frac{1}{n^2}\right)}
	.
\end{multline}
The vanishing of $A(0)$ in the low temperature phase ($\alpha < \alpha_{c}$)
has the consequence that the two limits $n \to \infty$ and $h \to 0$
do not commute, indeed:
\begin{align}
	& \lim_{n \to \infty} \lim_{h \to 0} m(\alpha,h) =
	\left. - \frac{\de B(h) }{ \de h}\,\right|_{h=0} - \frac{1}{C(0)} \left. \frac{\de A(h) }{ \de h}\,\right|_{h=0} = 0 \\
	& \lim_{h \to 0} \lim_{n \to \infty} m(\alpha,h) =
	\left. - \frac{\de B(h) }{ \de h}\,\right|_{h=0} = \frac{\alpha_{c} - \alpha}{\alpha_{c}} \geq 0
	.
\end{align}
Remark that the magnetization $m$ vanishes at the critical point linearly and not with critical exponent $1/2$ as it is common in mean-field theory, the reason being that here the order parameter is not linear but quadratic in the fundamental fields.

In the high temperature phase $A(0) \neq 0$ and the two limits above commute.
\begin{equation}
	\lim_{n \to \infty} \lim_{h \to 0} m(\alpha,h) =
	\lim_{h \to 0} \lim_{n \to \infty} m(\alpha,h) =
	\left. - \frac{\de B(h) }{ \de h}\,\right|_{h=0} = 0
	.
\end{equation}

In the study of phase transitions the thermodynamical limit $n
\to \infty$ has to be taken first. Indeed, the ergodicity is broken in the
thermodynamical limit first and then a residual spontaneous magnetization
appears even when the external field vanishes. Remark that both the free energy and the magnetization vary
continuously passing from one phase to the other.

The longitudinal susceptibility $\chi_L$ 
\begin{align}
	\chi_L(\alpha,h) = \, &  \frac{\de^2 s (\alpha,h) }{ \de h^2} \\
	= \, & 
	 \alpha \, (\alpha n -1) \, \frac{ \langle (\bar\psi \psi)^2 \,\left( {\cal U} + h \bar\psi \psi\right)^{\alpha n -2}\rangle_{t=0}}{\langle \left( {\cal U} + h \bar\psi \psi\right)^{\alpha n }\rangle_{t=0}}  - n \,\left[1 - m(\alpha,h)\right]^2
\end{align}
can be obtained from the magnetization:
\begin{multline}
	\chi_L(\alpha,h) = - \frac{\de m (\alpha,h) }{ \de h} =
		\frac{\de^{2} B(h) }{ \de h^{2}}
	+ \frac{1}{2n} \frac{1}{B''(h)^{2}} \left(\frac{\de B''(h) }{ \de h}\right)^{2}
	\\
	- \frac{1}{2n} \frac{1}{B''(h)} \frac{\de^{2} B''(h) }{ \de h^{2}}
	+ \frac{1}{n}
	\frac
	{\frac{\de^{2} A(h) }{ \de h^{2}} + \frac1n \frac{\de^{2} C(h) }{ \de h^{2}} + O(\frac1{n^2})}
	{ A(h) + \frac1n C(h) + O(\frac1{n^2})}
	\\
	- \frac{1}{n}
	\left[
	\frac
	{\frac{\de A(h) }{ \de h} + \frac1n \frac{\de C(h) }{ \de h} + O(\frac1{n^2})}
	{ A(h) + \frac1n C(h) + O(\frac1{n^2})}
	\right]^{2}
	+ O\left(\frac1{n^2}\right)
	,
\end{multline}
and taking the two limits in the appropriate order we get
\begin{align}
	\lim_{h \to 0} \lim_{n \to \infty} \chi_L(\alpha,h) =
	\left. \frac{\de^{2} B(h) }{ \de h^{2}}  \,\right|_{h=0} =
	\begin{cases}
		- \frac{\alpha(1 - \alpha)}{\alpha_{c}^{2}}
		\left( \frac{\alpha_{c}- \alpha}{\alpha_{c}} \right)^{-1} & \alpha < \alpha_{c} \\
		- \frac{1- \alpha_{c}}{\alpha_{c}}
		\left( \frac{\alpha - \alpha_{c}}{\alpha_{c}} \right)^{-1} & \alpha > \alpha_{c} \\
	\end{cases}
\end{align}
which shows that the susceptibility is discontinuous at the transition, with a singularity
$\chi(\alpha) \sim |\alpha-\alpha_c|^{-1}$, so that the transition is second order.
Remark that the longitudinal susceptibility appears to be negative. This means that in our model of spanning hyperforest there are events negatively correlated. 
It is well known that in the model of spanning trees on a finite connected graph the indicator functions for the events in which an edge belongs to the tree are negatively correlated. 
This is proven by Feder and Mihail~\cite{FM} in the
wider context of balanced matroids (and uniform weights). See also~\cite{40new} for a purely combinatorial proof of the stronger Raileigh
condition, in the weighted case. The random cluster model for $q>1$ is known to be positive associated. When $q<1$ negative association is conjectured to hold.
For an excellent description of the situation about negative association see~\cite{Pemantle}.

Still following the analogy with magnetic systems, let us introduce the transverse susceptibility
\begin{equation}
\chi_T(\alpha, h) :=  \frac{2}{\calZ_{\alpha n} }\,\frac{\< (\bar\psi, \J \psi)\, \calU^{\alpha n-1}\>_{t=0}}{n\, \Gamma(\alpha n)}
\end{equation}
which, by comparison with~(\ref{eq:t-square-saddle}), provides, at $h=0$
\begin{equation}
\chi_T(\alpha, 0) = \frac{2}{n} \, \<\, |T|^2\>_{\alpha n} \, .
\end{equation}	
In Appendix~\ref{sec:b} we prove the identity
\begin{equation}
m(\alpha, h) = \frac{h}{2}\, \chi_T(\alpha, h)\, .
\end{equation}	
This relation is the bridge between the average square-size of hypertrees and the local order parameter.

At finite $n$, when the symmetry-breaking field $h$ is set to zero, we get
\begin{equation}
m(\alpha, 0) = 0
\end{equation}	
in agreement with formula~(\ref{W1}) and
\begin{equation}
\chi_T(\alpha, 0) = \lim_{h\to 0} 2\,\frac{m(\alpha, h) }{h}  = \left. 2\,\frac{\partial m(\alpha, h) }{\partial h}\right|_{h=0} = - 2\,\chi_L(\alpha, 0)
\end{equation}	
which should be compared with the analogous formula for the $O(N)$-model where it is
\begin{equation}
\chi_T(\alpha, 0) = (N-1)\, \chi_L(\alpha, 0)
\end{equation}	
and, in our case, as the symmetry is $\osp(1|2)$, $N$ should be set to $-1$ as we have one bosonic direction and two fermionic ones which give a negative contribution.

The leading $n$ contribution is
\begin{equation}
\frac{1}{2}\,\chi_T(\alpha, 0) = - \, \chi_L(\alpha, 0) = \begin{cases}
n\,\left( \frac{\alpha_c - \alpha}{\alpha_c}\right)^2 & \hbox{for\, } \alpha \leq \alpha_c \\
 \frac{1}{k}\,  \frac{1}{\alpha- \alpha_c} & \hbox{for\, } \alpha \geq \alpha_c \, 
\end{cases}
\end{equation}	
But, for $ \alpha \leq \alpha_c$, if we first compute the large $n$ limit and afterwards send  $h \to 0$, we know that we get a non-zero magnetization and therefore the transverse susceptibility diverges as
\begin{equation}
\chi_T(\alpha, 0) \sim 2\, \frac{m(\alpha, 0)}{h} = \frac{2}{h}\, \frac{\alpha_c - \alpha}{\alpha_c}
\end{equation}	
which corresponds to the idea that there are massless excitations, Goldstone modes associated to the symmetry breaking.
Remark that, at finite $h$, the transverse susceptibility does not increase with $n$, which shows that the average square-size of hypertrees stays finite.

The longitudinal susceptibility instead
\begin{equation}
\chi_L(\alpha, 0) \sim  \, - \frac{1}{m(\alpha,0)}\,\frac{\alpha\,(1-\alpha)}{\alpha_c^2}\, .
\end{equation}	
diverges only at $\alpha = \alpha_c$, when the magnetization vanishes.


\section{A symmetric average}
\label{sec:c}

At the breaking of an ordinary symmetry the equilibrium states can be written as a convex superposition of pure, clustering, states, which cab be obtained, one from the other, by applying the broken symmetry transformations. The pure state we have defined in this Section uses a breaking field in the only direction we have at disposal where the Grassmann components are null. A more general breaking field would involve a direction in the superspace to which we are unable to give a combinatorial meaning. However, if we take the average in the invariant Berezin integral of these fields we give rise to a different, non-pure but symmetric, low-temperature state. 

In this Section we shall set $t=1$.

The most general breaking field, with total strength $h$, but arbitrary direction in the super-space, would give a weight
\begin{equation}
h\, \sum_{i=1}^n \left[  \lambda\,( 1 - \bar\psi_i \psi_i) +\, \bar \epsilon \psi_i + \, \bar \psi_i \epsilon\,\right]
\end{equation}
where $(\lambda; \bar{\epsilon}, \epsilon)$ is a unit vector in the
$1|2$ supersphere, i.e.~$\epsilon$ and $\bar{\epsilon}$ are Grassmann
coordinates and $\lambda$ is a formal variable satisfying the
constraint $$ \lambda^2 + 2 \, \bar \epsilon \, \epsilon = 1 .$$
Let us introduce the normalized  generalized measure
\begin{equation}
d\Omega := d\lambda\, d\epsilon \, d\bar \epsilon  \, \delta\left( \lambda^2 + 2 \, \bar \epsilon \, \epsilon - 1\right)\, .
\end{equation}
A symmetric equilibrium measure can be constructed by considering the factor
\begin{align}
& F[h; \bar\psi, \psi] := \\
& = \int d\Omega\, \exp \left\{  \, -\, h\,\sum_{i=1}^n \left[  \lambda\,( 1 - \bar\psi_i \psi_i) +\, \bar \epsilon \psi_i + \, \bar \psi_i \epsilon\,\right]\right\} \\
& = \int  d\epsilon \, d\bar \epsilon \, \exp \left\{  \bar \epsilon\, \epsilon  -   h\, \sum_{i=1}^n \left[ (1 -  \bar \epsilon\, \epsilon) ( 1 - \bar\psi_i \psi_i) + \bar \epsilon \psi_i + \bar \psi_i \epsilon\right]\right\}\, \\
& = \left[ 1 - h^2\, ( \bar\psi, \J  \psi) + h\,(n -  \bar \psi \psi) \right]\, \exp \left[ -\,h\,(n -  \bar \psi \psi)  \right]. 
\end{align}
where only the last expression is specific to our model, but the previous are the appropriate expressions for the model of unrooted spanning hyperforests on an arbitrary weighted hypergraph.
This function is symmetric, for every strength $h$, as it can be easily checked that
\begin{equation}
Q_\pm\,  F= 0 \, .
\end{equation}

If we send $h\to 0$ this factor is simply $1$, but if we first take the $n\to \infty$ limit and then $h\to 0$ the expectation value of non-symmetric observables can be different.

The partition function is not changed because of the identity~(\ref{aw1}). Indeed 
\begin{equation}
\< F \> = \< \exp \left[ -\,h\,(n -  \bar \psi \psi)  \right] \> = \< 1  \>_h
\end{equation}
for un-normalized expectation values, because of the relation between the transverse susceptibility and the magnetization, equation~(\ref{aw1}), which is
\begin{align}
0 = &\, \< \left[- h^2\, ( \bar\psi, \J  \psi) + h\,(n -  \bar \psi \psi) \right]\, \exp \left[ -\,h\,(n -  \bar \psi \psi)  \right] \>  \\
= &\, h\, \< \left[- h\, ( \bar\psi, \J  \psi) + \,(n -  \bar \psi \psi) \right] \>_h \, 
\end{align}
for every $h$, and therefore also for the derivatives with respect to $h$.
But consider for example the magnetization. The insertion of the given factor $F$ in the un-normalized expectation provides the relation
\begin{align}
& \< (n -  \bar \psi \psi)\>^{\mathrm{sym}}_h :=   \< (n -  \bar \psi \psi)\, F\>  = \< (n -  \bar \psi \psi)\>_h + \\
&  \; +   \<\left[  - h^2\, ( \bar\psi, \J  \psi) + h\,(n -  \bar \psi \psi) \right]\,\left( - \frac{\partial}{\partial h} \right) \exp \left[ -\,h\,(n -  \bar \psi \psi)  \right]  \> \\
& = \, 2\, \< (n -  \bar \psi \psi)\>_h - 2\, h\, \<( \bar\psi, \J  \psi)\>_h = 0 \,
\end{align}
Similarly
\begin{align}
 \< ( \bar\psi, \J  \psi)\>^{\mathrm{sym}}_h  & =  \<( \bar\psi, \J  \psi)\, F \> \\
& =  \<( \bar\psi, \J  \psi) \>_h - h  \frac{\partial}{\partial h}  \<( \bar\psi, \J  \psi) \>_h \\
& =  \<( \bar\psi, \J  \psi) \>_h - h     \frac{\partial}{\partial h}   \frac{\< (n -  \bar \psi \psi)\>_h }{h} \\
& =  \<( \bar\psi, \J  \psi) \>_h  + \left( \frac{1 }{h}  -   \frac{\partial}{\partial h} \right) \,\< (n -  \bar \psi \psi)\>_h  \\
& = 2\, \<( \bar\psi, \J  \psi) \>_h +  \< (n -  \bar \psi \psi)^2\>_h \\
& = \sum_{i,j} \,\<  \bar\psi_i \psi_j +  \bar\psi_j \psi_i + (1 -  \bar \psi_i \psi_i)(1 -  \bar \psi_j \psi_j)\>_h \\
& = \sum_{i,j} \,\< 1 - f_{\{i,j\}} \>_h
\end{align}
is the {\em total}, not-connected, susceptibility, that is the sum of the longitudinal and transverse not-connected ones. And also
\begin{equation}
 \< (n -  \bar \psi \psi)^2 \>^{\mathrm{sym}}_h   =  - \,2\, \<( \bar\psi, \J  \psi) \>_h -  \< (n -  \bar \psi \psi)^2\>_h 
\end{equation}
so that
\begin{equation}
2\, \<( \bar\psi, \J  \psi) \>^{\mathrm{sym}}_h +  \< (n -  \bar \psi \psi)^2\>^{\mathrm{sym}}_h =  2\, \<( \bar\psi, \J  \psi) \>_h +  \< (n -  \bar \psi \psi)^2\>_h
\end{equation} 
as it must occur for a symmetric observable.

\section{Conclusions} 
\label{sec:conclusions}
We have found that in the $k$-uniform complete hypergraph with $n$ vertices, in the limit of large $n$,  the structure of the hyperforests with $p$ hypertrees has an abrupt change when $p = \alpha_c n$ with $\alpha_c = (k-1)/k$. 
This change of behaviour is related to the appearance of a {\em giant} hypertree which covers a finite fraction of all the vertices.
As the number of hyperedges in the hyperforests with $p$ hypertrees is $(n-p)/(k-1)$, this means that this change occurs when the number of hyperedges becomes $1/k(k-1)$, which is exactly the critical number of hyperdges in the phase transition of random hypergraphs at fixed number of hyperedges~\cite{SPS}.

If ${\cal Z}(t)$ is the generating partition function of hyperforests, where the coefficient of  $t^p$ is the total number of those with $p$ hypertrees, in the limit of large $n$ there is a corresponding singularity at $t_c = n\, [(k-2)!]^{-1/(k-1)}$.

In our Grassmann formulation this singularity can be described as a second-order phase transition associated to the breaking of a global $\osp(1|2)$ supersymmetry which is non-linearly realised.
The equilibrium state occurring in the broken phase can be studied by the introduction of an explicit breaking of the supersymmetry.


\appendix 

\section{Saddle-point constants } 
\label{sec:appendix}

Let us use the notation
$$
X_{(n)} := \left. \frac{\partial^n}{\partial \eta^n} X(\eta) \right |_{\eta = \eta^*}\, .
$$ 
For the simple saddle point we report the combinations $C$ and $D$ in terms of functions
$A$ and $B$
\begin{align}
\label{eq:single-saddle-c-and-d}
	C & = 
	\frac{1}{24 B_{(2)}^3} \left[
		12 A_{(1)} B_{(2)} B_{(3)}
		- 12 A_{(2)} B_{(2)}^2
		+ A \left(
			3 B_{(2)} B_{(4)} 
			- 5 B_{(3)} ^2
		\right)
	\right]
	\\
    D & = \frac{1}{1152 B_{(2)}^6}
    \left[
    		385\, A \,B_{(3)}^4
		+ 144\, B_{(2)}^4 A_{(4)}
		\right.  \nonumber \\ & \quad \left.
		- 210\, B_{(2)} B_{(3)}^2 \left(
			4 A_{(1)} B_{(3)}
			+ 3 A B_{(4)}
      \right)
    \right. \nonumber \\ & \quad \left.
    + 21\, B_{(2)}^2 \left(
    		40 A_{(2)} B_{(3)}^2
		+ 40 A_{(1)} B_{(3)} B_{(4)}
		+ 5 A B_{(4)}^2
		+ 8 A B_{(3)} B_{(5)}
    \right) \right. \nonumber \\ & \quad \left.
    - 24\, B_{(2)}^3 \left(
    		20 A_{(3)} B_{(3)}
		+ 15 A_{(2)} B_{(4)}
		+ 6 A_{(1)} B_{(5)}
		+ A B_{(6)}
	\right) \right] \, .
\end{align}
For the double saddle points the necessary combinations are instead
\begin{align}
	\label{eq:double-saddle-c-and-d}
	\tilde C &= \frac
	{A \, B_{(4)}}
	{B_{(3)}^{4/3}}
	\frac{\gamma_{4}}{4!}
	- \frac{A_{(1)}}{B_{(3)}^{1/3}}
	\gamma_{1} \\
	\tilde D &=
	- \frac{A_{(3)}}{B_{(3)}} \frac{\gamma_{3}}{3!}
	+ \frac{A_{(2)} B_{(4)}} {B_{(3)}^{2}}
	\frac{\gamma_{6}}{2 \cdot 4!}
	+ \frac{A_{(1)} B_{(5)}}{B_{(3)}^{2}}
	\frac{\gamma_{6}}{5!}
         - \frac{A_{(1)} B_{(4)}^{2}}{B_{(3)}^{3}}
	\frac{\gamma_{9}}{2 (4!)^{2}}
	\nonumber \\ & \quad
	- \frac{A \, B_{(4)} B_{(5)}}{B_{(3)}^{3}}
	\frac{\gamma_{9}}{4!5!}
	+ \frac{A \, B_{(4)}^{3}}{B_{(3)}^{4}}
	\frac{\gamma_{12}}{3! (4!)^{3}}
	+ \frac{A \, B_{(6)}}{B_{(3)}^{2}}
	\frac{\gamma_{6}}{6!} \, .
\end{align}
The constants $\gamma_{k}$ are given by
\begin{align}
	\gamma_{k}   :=
	& -\frac{1}{\pi}
\sin \left( 2 \pi \, \frac{1+k}{3} \right)
\int_0^{\infty} du\, u^k \, e^{-\frac{u^3}{3!}} \\  = 	
	& - \frac{(3!)^{\frac{1 + k}{3}}}{3\pi} \sin\left[2 \pi \left(\frac{1 + k}{3}\right)\right]
	\Gamma\left(\frac{1+k}{3}\right)\, .
\end{align}


\section{Ward identities}
\label{sec:b}

As a result of the underlying symmetry, there are relations among the correlation functions, called {\em Ward identities}~\cite{Zinn-Justin}.
In this Appendix we give a more direct derivation of one of them which simply uses integration by parts.

By definition
\begin{equation}
{\cal U}(\xi) \, = \, \xi + (1 - k) \frac {\xi^k} {k!} 
\end{equation}
so that
\begin{equation}
 \frac{\partial {\cal U}} {\partial \xi} =  1 - \frac {\xi^{k-1}}{(k-2)!}
 \end{equation}
and therefore the un-normalized expectation value of $\bar\psi \psi$ in presence of the symmetry breaking is
\begin{align}
&  t\, \< \bar\psi \psi \>_h \, = \nonumber \\
&	=  \, n! \oint \frac {d\xi} {2 \pi i} \, \frac1{\xi^{n+1}}\, \exp \left\{t\, {\cal U} + n \frac {\xi^{k-1}} {(k-1)!} - h\, t \, (n - \xi)  \right\} \, t\,\xi\, \frac{\partial {\cal U}} {\partial \xi} \\
&	=  \, n! \oint \frac {d\xi} {2 \pi i} \, \frac1{\xi^{n+1}}\, \exp \left\{ n \frac {\xi^{k-1}} {(k-1)!}  - h\, t \, (n - \xi) \right\} \,\xi\, \frac{\partial} {\partial \xi} \,\exp \left\{t\, {\cal U} \right\}\, .
\end{align}
Perform now an integration by parts
\begin{align}
&  t\, \< \bar\psi \psi \>_h \, = \nonumber \\
&	=  \, {n!} \oint \frac {d\xi} {2 \pi i} \, \frac1{\xi^{n+1}}\, \left[ {n} \left( 1 -  \frac{\xi^{k-1}}{(k-2)!} \right)   - h\, t\,\xi \right] \,  \nonumber \\
& \qquad \qquad \qquad \qquad \qquad  \exp \left\{t\, {\cal U} + n \frac {\xi^{k-1}} {(k-1)!}  - h\, t \, (n - \xi) \right\} \\
 & = \,n\, {\cal Z}(t, h)\, - \, h\, t\, \< (\bar\psi, \J \psi) \>_h\, . 
\end{align}
So that
\begin{equation}
\,n\, {\cal Z}(t, h)\, - \, t\, \< \bar\psi \psi \> \, =  \, h\, t\, \< (\bar\psi, \J \psi) \>\,  \label{aw1}
\end{equation}
which expanded in series of $t$ implies that
\begin{multline}
\,\frac{n}{p}\, \<  ({\cal U} + h  \bar\psi \psi)^{p}\>_{t=0} \ - \,  \< \bar\psi \psi\, ({\cal U} + h  \bar\psi \psi)^{p-1}\>_{t=0} \, =  \, \\
 h\, \< (\bar\psi, \J \psi) ({\cal U} + h  \bar\psi \psi)^{p-1} \>_{t=0}\,  
\end{multline}
or for $p=\alpha n$
\begin{multline}
1 \ - \, \alpha\, \frac{ \< \bar\psi \psi\, ({\cal U} + h  \bar\psi \psi)^{\alpha n-1}\>_{t=0} }{ \<  ({\cal U} + h  \bar\psi \psi)^{\alpha n}\>_{t=0}}\, =  \, \\
\alpha\,  h\,\frac{ \< (\bar\psi, \J \psi) ({\cal U} + h  \bar\psi \psi)^{\alpha n-1} \>_{t=0} }{ \<  ({\cal U} + h  \bar\psi \psi)^{\alpha n}\>_{t=0}}\,  \label{aw1m}
\end{multline}
which means that, in the microcanonical ensemble, for every $h$, we have
\begin{equation}
m(\alpha, h) \, = \, \frac{h}{2}\, \chi_T(\alpha, h)\, . \label{ward}
\end{equation}

\end{document}